\documentclass[aps,prl,twocolumn,showpacs,groupedaddress,amsmath,floatfix,11pt,reprint]{revtex4-2} 

\usepackage{graphicx}  
\usepackage{multirow}
\usepackage{makecell}
\usepackage{dcolumn}   
\usepackage{bm}        
\usepackage{amssymb}   
\usepackage{framed}
\usepackage{amsmath}
\usepackage{hhline}
\usepackage{xcolor}
\usepackage{subfigure,amsmath,verbatim,moreverb}
\usepackage{tabularx}
\usepackage{lipsum}
\usepackage{longtable}
\usepackage{booktabs}
\usepackage{adjustbox}
\usepackage{braket}
\usepackage{enumitem}
\usepackage{etoolbox}
\AtBeginEnvironment{align}{\setcounter{subeqn}{0}}
\newcounter{subeqn} 

\newcommand{\R}{\mathbf{r}}

\begin{document}

\begin{abstract}
We develop a non-linear and non-empirical (nlane) double hybrid density functional derived from an accurate interpolation of the adiabatic connection in density functional theory, incorporating the correct asymptotic expansions. 
By bridging the second-order perturbative weak correlation limit with the fully interacting limit from the semi-local SCAN functional, nlane-SCAN is free of fitted parameters while providing
improved energetic predictions compared to SCAN for moderately and strongly correlated systems alike. 
It delivers accurate predictions for atomic total energies and multiple reaction datasets from the GMTKN55 benchmark while significantly outperforming traditional linear hybrids and double hybrids for non-covalent interactions without requiring dispersion corrections. 
Due to the exact constraints at the weak correlation limit, nlane-SCAN has reduced delocalization errors as evident through the SIE4x4 benchmark and bond dissociations of H\(_2^+\) and He\(_2^+\). 
Its proper asymptotic behavior ensures stability in strongly correlated systems, improving H\(_2\) and N\(_2\) bond dissociation profiles compared to conventional functionals.
\end{abstract}

\title{Non-linear and non-empirical double hybrid density functional}
\author{Danish Khan}
\thanks{\texttt{danishk.khan@mail.utoronto.ca}}
\affiliation{Chemical Physics Theory Group, Department of Chemistry, University of Toronto,\\ St. George Campus, Toronto, ON, Canada}
\affiliation{Vector Institute for Artificial Intelligence, Toronto, ON, M5S 1M1, Canada}

\maketitle
\section{Introduction}
Kohn-Sham density functional theory (KS-DFT) is the workhorse of computational simulations in chemistry and materials science due to its standout compute cost vs accuracy tradeoff~\cite{burke12,perdew2024my}.
Its importance has grown further in the artificial intelligence age since virtually all reference data used by machine learning models comes from KS-DFT~\cite{central_role_dft}.
At the heart of KS-DFT lie the exchange-correlation (XC) functionals, which approximate the only unknown part of the KS energy functional~\cite{mattsson2002pursuit} and allow defining suitable density functional approximations (DFAs) with varying levels of computational accuracy and cost.
These KS-orbital/electron density functionals come in two broad flavours, semi-local and non-local, depending on whether the XC-potential at each grid point in space relies on information present locally or not~\cite{becke14}.
Due to their greater computational efficiency, semi-local functionals largely underpin the popularity and success of KS-DFT, especially in materials~\cite{jones2015density}.

While impressively accurate for a wide range of tasks, all semi-local functionals inadvertently suffer from self-interaction error (SIE) caused delocalization errors~\cite{bryenton2023delocalization}.
Self-interaction, or delocalization errors in general, manifest in a number
of well known ways, from the poor performance on band gaps~\cite{cohen08} to inaccurate prediciton of several chemically relevant systems, such as charge-transfer complexes, charge transfer excitation energies, halogen bonded complexes, molecular crystals as well as barrier heights of radicals in general~\cite{bryenton2023delocalization}.
Delocalization errors can be significantly mitigated through incorporation of wavefunction theory based non-local functionals of the KS-orbitals in the form of Hartree-Fock (HF) exchange and second order Møller–Plesset (MP2) correlation energies within hybrid, double-hybrid functionals respectively.
This "fix", however, has a couple of key shortcomings, the primary of which is a "zero-sum" tradeoff~\cite{janesko2021replacing, mori2014derivative} between mitigating delocalization errors, and capturing static correlation.
In particular, while double-hybrid functionals achieve remarkable accuracy for covalently bonded systems and even reaction barrier heights~\cite{goerigk2014double}, they fail catastrophically for systems with strong static correlation wherein the MP2 correlation energy diverges~\cite{bochevarov2005hybrid}.
This happens since systems with strong static correlation consist of multiple near-degenerate Slater determinants in configuration space, which shrinks the difference in the frontier orbital eigenvalues to near 0 when computed through a single-determinantal method.
Consequently, the MP2 correlation energy (eq.~\ref{Ec_mp2}) diverges.
Traditional double-hybrid functionals mix MP2 correlation energy linearly with the DFA energy and thereby inherit this divergent behaviour. 
This is easy to see through simple closed-shell bond-dissociations where these functionals perform even worse than semi-local functionals~\cite{hui2016scan}.
The second key shortcoming of these functionals is the arbitrary choice of the HF exchange and MP2 correlation mixing fractions which are obtained through empirical fitting.
While fitting to empirical data optimizes their performance within a specific subset of the chemical space, extrapolation is not guaranteed.

Both of these shortcomings are fixed by functionals derived through the adiabatic connection (AC) formalism \cite{langreth75,gunnarsson76,savin03}, a general, powerful tool for the development of XC functionals. 
For several decades it has been used to justify the introduction of hybrid \cite{becke93,pbe0,fabiano15} and double hybrid functionals \cite{sharkas11,bremond11,bremond16} and successively it has been directly employed to construct high-level XC functionals based on AC models (ACM) interpolating between known limits of the exact AC integrand (eq.~\ref{adiabatic_connection}) \cite{isi,seidl00,liu09,ernzrehof99} .
Functionals derived in such a way do not generally require any fitted parameters, and can incorporate the strong correlation limit of the AC, thereby significantly improving description of these difficult cases~\cite{teale10}.
Unfortunately, such functionals are inapplicable to most equilibrium cases where traditional non-local, as well as semi-local, functionals excel~\cite{goerigk2017look} since they can be highly inaccurate for the description of covalently bonded systems~\cite{isiHF_analysis2016}.
The primary reason behind this is the lack of known constraints of the AC integrand for the physical, $\lambda=1$, system which is not accounted for within the construction of such functionals.
For most equilibrium systems with negligible static correlation (or large HOMO-LUMO gaps), the $\lambda \xrightarrow[]{} \infty$ limit of the AC within DFT~\cite{seidl00} is irrelevant~\cite{gorigiorgi09}.

Within this work we will build upon the aforementioned attractive features of ACM functionals, while replacing the explicit strong-correlation limit with an estimate of the $\lambda=1$ limit using the accurate semi-local SCAN functional~\cite{sun2015strongly}.
Since we will retain the exact weak-coupling limit, HF exchange and MP2 correlation energies will enter our double-hybrid functional naturally through exact constraints thereby not requiring any parameter fitting.
Consequently, our double-hybrid functional will have significantly reduced delocalization errors and will be free from one-electron SIE entirely, as shown through the SIE4x4 benchmark~\cite{SIE4x4} and bond-dissociations of H\(_2^+\) and He\(_2^+\).
The strong-coupling limit~\cite{seidl00} will be incorporated implicitly in the chosen form of the AC integrand which will ensure correct behaviour for strongly correlated systems, free from any divergence as confirmed through bond-dissociations of H\(_2\) and N\(_2\).
Furthermore, since we explicitly incorporate an approximation to the physical $\lambda=1$ limit, the performance of our functional for general thermochemistry and kinetics of covalently bounded systems is not hindered.
Instead, it significantly improves upon the semi-local SCAN DFA used for this approximation while also outperforming empirical linear hybrids and double-hybrids for multiple datasets from the GMTKN55 benchmark~\cite{goerigk2017look}.
Lastly, we will also discuss future avenues for improving the $\lambda=1$ approximation through adaptive machine learning schemes as developed for in ref.~\cite{khan2025adapting} for the PBE0~\cite{pbe0} functional.

\section{Theory and methods}
\begin{figure}[!ht]
    \centering
    \includegraphics[width=\linewidth]{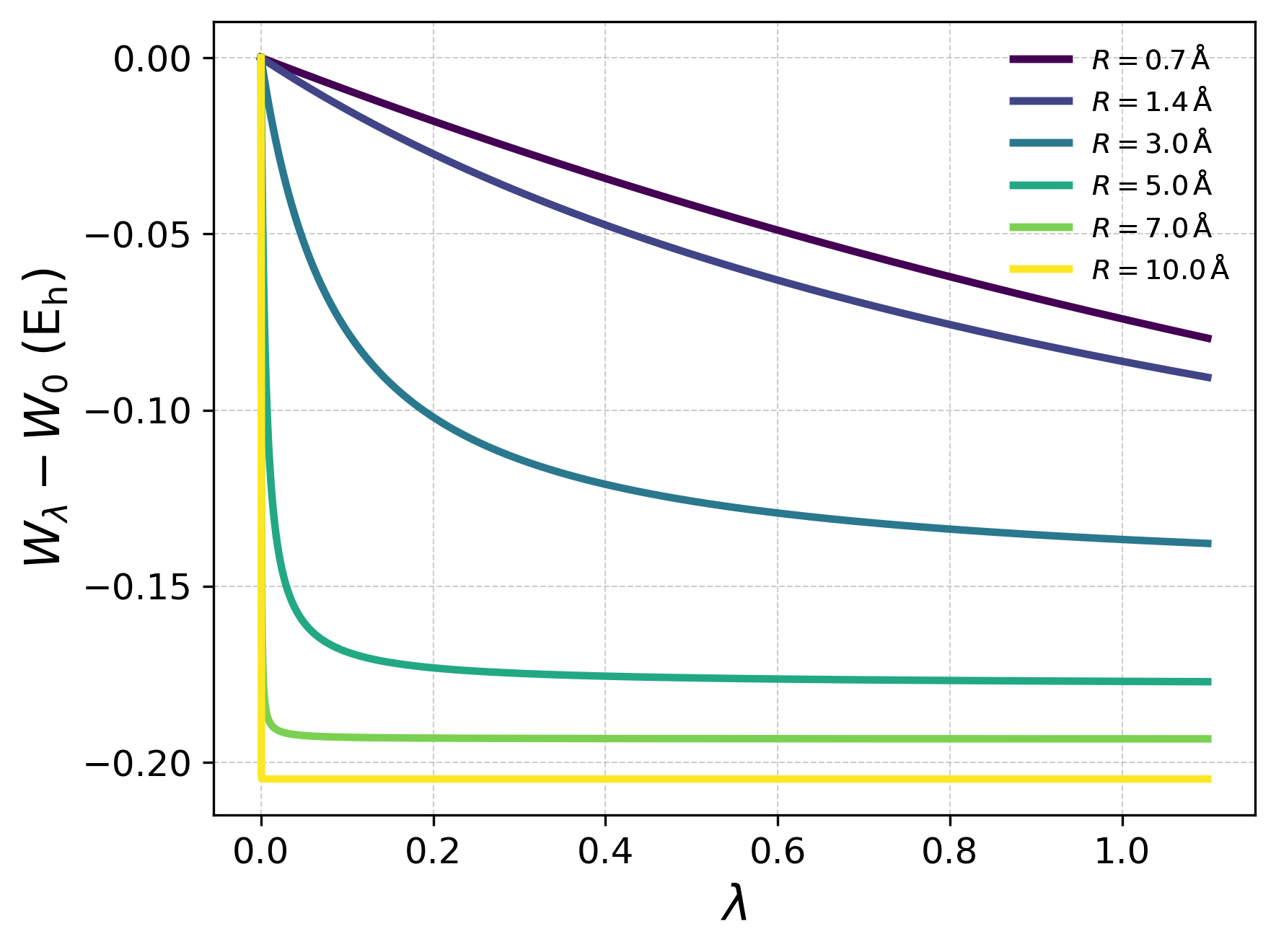}
    \caption{nlane-SCAN's (correlation) adiabatic connection curves for the H\(_2\) molecule at different bond-lengths $R$.
    The curves plot eq.~\ref{ansatz_AC} with the $W_{0}, W'_{0}, W_{1}$ values from eqs.~\ref{constraint1}, \ref{constraint2}, \ref{constraint3} as constraints for obtaining $a,b,c$.
    See figure 8 in Ref.~\cite{teale10} for the corresponding exact (Full-CI) adiabatic connection curves.
    }
    \label{fig:h2_ac}
\end{figure}

XC functionals based on AC models (ACM) interpolate between known limits of the exact AC integrand \cite{isi,seidl00,gorigiorgi09,liu09,ernzrehof99,teale10} $W_{\lambda}$
\begin{align}
    W_{\lambda} = \braket{\Psi_{\lambda}|\hat{v}_{ee}|\Psi_{\lambda}} - J[\rho]
    \label{adiabatic_connection}
\end{align}
Here $\rho$ denotes the ground state density of the physical system, $\Psi_{\lambda}$ is the ground-state wavefunction of the intermediate system with inter-electronic interactions scaled by $\lambda$ and same density as $\rho$, $\hat{v}_{ee}$ denotes the inter-electronic repulsion operator and $J$ denotes the Coulomb energy $J[\rho] = \frac{1}{2}\int \int \rho(\mathbf{r}_{1}) \rho(\mathbf{r}_{2}) r^{-1}_{12} d\mathbf{r}_{1} d\mathbf{r}_{2}$ of the density $\rho$.
The XC functionals based on ACMs, employing a model $W_\lambda^{ACM}$, are then derived from the integral over $\lambda$
\begin{equation}\label{e1}
E_{\rm xc}^{ACM} = f^{ACM}(\mathbf{W}) = \int_0^1W_\lambda^{ACM}(\mathbf{W})d\lambda\ 
\end{equation}
where $\mathbf{W}=(W_0,W'_0,W_\infty,W'_\infty)$, with
$W_0=E_{\rm x}^{\rm HF}$ being the Hartree-Fock (exact) exchange energy on the Kohn-Sham (KS) orbitals,
$W'_0=2E_{\rm c}^{\rm GL2}$ being twice the correlation energy from second-order Görling-Levy (GL) perturbation theory~\cite{gl2}, and $W_\infty$ and $W'_\infty$
being the indirect part of the minimum expectation value
of the electron-electron repulsion for a given density and the potential energy of coupled zero-point oscillations around this minimum, respectively \cite{seidl07,gorigiorgi09}.
The model $W_\lambda^{ACM}$ is designed to mimic the exact but unknown $W_\lambda$, in particular by considering the known asymptotic expansions \cite{gorigiorgi09,SeiPerLev-PRA-99,gl2,seidl07}
\begin{eqnarray}
W_{\lambda\rightarrow 0} & = & E_{\rm x}^{\rm HF} + \sum_{n=2}^{\infty} n E_{\rm c}^{\rm GLn} \lambda^{n-1} \label{GLn} \\
W_{\lambda\rightarrow \infty} & = & W_\infty + W'_\infty \lambda^{-1/2} +W''_\infty \lambda^{-3/2}+ \mathcal{O}(\lambda^{-5/2})\ \label{asymptote}
\end{eqnarray}
and by incorporating the known properties of the $W_{\lambda}$ path~\cite{AC_review_DAAS_Vuckovic}.
Functionals constructed in such a way perform remarably well for strongly correlated systems but can be rather inaccurate for general main-group thermochemistry and kinetics~\cite{fabiano2016ISI_limitations} where standard density functional approximations (DFAs) excel.
\\
To overcome this issue, in this work we will derive a functional that incorporates information from the fully interacting physical limit of the AC, i.e. $\lambda=1$, rather than the $\lambda \rightarrow \infty$ limit.
This is similar in spirit to the construction attempted within the MCY functionals~\cite{mori2006many} aimed at minimizing self-interaction errors.
These functionals, however, did not have the correct $\lambda \rightarrow \infty$ asymptotic behaviour in eq.~\ref{asymptote}, used an empirically fitted estimate for the $\lambda = 1$ limit, and the exact slope at the weak-coupling limit was approximated through a semi-local DFA for computational efficiency.
These shortcomings are rectified within our construction which ensures the correct behaviour in strong correlation cases while retaining the robustness of double-hybrid DFAs for thermochemistry of regular, non-strongly correlated, systems.
Hence, the ingredients used within our ACM correspond to $\mathbf{W}=(W_0,W'_0,W_1)$ with the $\lambda \rightarrow \infty$ asymptotic behaviour from eq.~\eqref{asymptote}
being incorporated implicitly through the chosen form of the function $W_{\lambda}$.
For this we modify the [1,1] Padé form proposed by Ernzerhof~\cite{ernzerhof1996construction}, and used within the MCY functionals, to
\begin{align}
    W_{\lambda} = a + \frac{b \sqrt{\lambda + 1}}{c\lambda + 1}
    \label{ansatz_AC}
\end{align}
where $a$, $b$, $c$ are density/orbital dependent quantities to be evaluated through the constraints based on $\mathbf{W}=(W_0,W'_0,W_1)$.
See Figure.~\ref{fig:h2_ac} for an illustration of this function's behaviour as the system transitions from moderately to strongly/strictly correlated regime.
Eq.~\eqref{ansatz_AC} inherits the required smoothness and convexity properties (for a list of requirements for the AC interpolants, see ref.~\cite{AC_review_DAAS_Vuckovic}) of the original Padé form, while correcting the $\lambda \rightarrow \infty$ asymptotic expansion similar to eq.~\ref{asymptote}:
\begin{align}
    W_{\lambda\rightarrow \infty} = a + \frac{b}{c} \lambda^{-1/2} + \frac{b(c-2)}{2c^2} \lambda^{-3/2} + \mathcal{O}(\lambda^{-5/2})
    \label{asymptote_func}
\end{align}
Since eq.~\ref{ansatz_AC} is an analytical function, it has a convergent Taylor series at $\lambda=0$, in line with the GL expansion, which can be used to impose two exact constraints from the weak-coupling limit in eq.~\ref{GLn} for calculating $a,b,c$
\begin{align}
   W_{0} = E_{\rm x}^{\rm HF} \implies a + b := E_{\rm x}^{\rm HF}
   \label{constraint1}
\end{align}
\begin{align}
   W'_{0} = 2 E_{\rm c}^{\rm GL2} \approx 2 E_{\rm c}^{\rm MP2} \implies b(\frac{1}{2} - c) := 2 E_{\rm c}^{MP2}
   \label{constraint2}
\end{align}
where $ E_{\rm c}^{\rm MP2}$ denotes the correlation energy from second-order Møller–Plesset (MP) perturbation theory on the KS occupited (occ.) and virtual (virt.) orbitals with eigenvalues $\varepsilon$:
\begin{align}
    E_{\rm c}^{\rm MP2} = -\frac{1}{4} \sum_{ab}^{\rm virt.} \sum_{ij}^{\rm occ.} \frac{\left|\braket{ab|\hat{v}_{ee}|ij} - \braket{ab|\hat{v}_{ee}|ji}\right|^2}{\varepsilon_{a} + \varepsilon_{b} - \varepsilon_{i} - \varepsilon_{j}}
    \label{Ec_mp2}
\end{align}
Note that $W'_{0} = 0$ only holds for one-electron systems wherein the perturbation ($E_{\rm C}^{\rm MP2}$) is 0 since the HF energy is exact.
In this case we have two valid solutions for eq.~\ref{constraint2}, i.e. $b=0$ or $c = 1/2$.
Due to the lack of a third exact constraint within our functional (see eq.~\ref{constraint3} below), we choose the first solution which implies $E_{\rm XC} = W_{\lambda} = a = E_{\rm X}^{\rm HF}$ for all one-electron systems, as it should.
This renders our functional free from one-electron self-interaction error (SIE) by construction, similar to the MCY functionals.
\\
Eq.~\ref{constraint2} contains an error due to the approximation made by ignoring the single-excitation term within the GL2 energy
\begin{align}
    E_{\rm c}^{\rm GL2} = E_{\rm c}^{\rm MP2}-\sum_{a}^{\rm virt} \sum_{i}^{\rm occ} \frac{\left|\braket{a|\hat{v}_{x}^{\rm KS} - \hat{v}_{x}^{\rm HF}|i}\right|^2}{\varepsilon_{a}  - \varepsilon_{i}}
\end{align}
where $\hat{v}_{x}^{\rm KS}$, $\hat{v}_{x}^{\rm HF}$ denote the local and non-local KS, HF exchange operators respectively.
Note that conventional linear double hybrid functionals compensate for this error through empirical fitting~\cite{XYG3}, but this error remains uncorrected within our functional.
\\
The first two constraints (eqs.~\ref{constraint1},~\ref{constraint2}) can be generalized to match higher order Taylor expansion of the interpolating function (eq.~\ref{ansatz_AC}) with exact values from perturbation theory (eq.~\ref{GLn}) at $\lambda=0$.
For instance, an additive parameter $d$ in the denominator of eq.~\ref{ansatz_AC} would still preserve the correct asymptotic expansion (eq.~\ref{asymptote}) and this can be used to add an additional constraint incorporating the GL3 energy as
\begin{align}
    W''_{0} = 6E_{\rm c}^{\rm GL3}
\end{align}
In the current work we restrict ourselves to the MP2 term term in eq.~\ref{constraint2} for computational efficiency.
\\
For an alternative third constraint we incorporate information from the, unknown, fully-interacting $\lambda=1$ limit of the physical system where semi-local functionals are much more accurate as the XC-hole is localized near the electron~\cite{becke2014perspective}.
In our work we rely on the semi-local strongly constrained and appropriately normed (SCAN)~\cite{sun2015strongly} functional, which has been noted to be the most accurate semi-local DFA for general thermochemistry~\cite{goerigk2017look} thanks to its rigorous construction~\cite{kaplan2023predictive}.
This provides the last required equation for calculating $a,b,c$
\begin{align}   
W_{1} \approx W_{1}^{\text{SCAN}} \implies a + \frac{b \sqrt{2}}{c+1} = W_{1}^{\text{SCAN}}    
\label{constraint3} 
\end{align}
Note that SCAN is a non-empirical functional derived by imposing 17 exact constraints of the universal functional.
See the supplement of ref.~\cite{sun2015strongly} for the list of exact constraints used, and ref.~\cite{kaplan2023predictive} for a discussion of all known exact mathematical constraints.
We point out here that SCAN was "normed" to a few atomic and di-atomic systems in order to tighten the inequality constraints used in the construction.
The $W_{\lambda}$ value at any arbitrary scaling $\lambda$ can be calculated for a semi-local DFA using the Levy-Perdew scaling relation\cite{levy1985hellmann}
\begin{equation}
W_{\lambda}^{\rm DFA}[\rho] =  E^{\rm DFA}_{\rm x}[\rho] +  2 E^{\rm DFA}_{\rm c}[\rho_{1/\lambda}]\lambda  +\lambda^2  \partial_{\lambda} E^{\rm DFA}_{\rm c}[\rho_{1/\lambda}] 
\label{Levy-Perdew}
\end{equation}
with $\rho_{1/\lambda}(\R)=\lambda^{-3}\rho(\R/\lambda)$ being the coordinate-scaled\cite{seidl2000density,gorling1993correlation} density.
This form simplifies at the physical limit $\lambda=1$
\begin{equation}
W_{1}^{\rm DFA}[\rho] =  E^{\rm DFA}_{\rm x}[\rho] +  2 E^{\rm DFA}_{\rm c}[\rho]  + \partial_{\lambda} E^{\rm DFA}_{\rm c}[\rho_{1/\lambda}]  \vert_{\lambda=1}
\label{Levy_perdew1}
\end{equation}
For simplicity, in our work we ignore the derivative term in this expression (this holds in the low-density limit) and use the following approximation in the third constraint (eq.~\ref{constraint3})
\begin{align}
    W_{1} \approx W_{1}^{\text{SCAN}} \approx E^{\rm SCAN}_{\rm x} +  2 E^{\rm SCAN}_{\rm c}
    \label{w1_scan}
\end{align}
with $E^{\rm SCAN}_{\rm x}, E^{\rm SCAN}_{\rm c}$ denoting the exchange and correlation energies from the SCAN functional respectively.
This approximation is another source of error within our functional which will be analyzed/fixed through automatic differentiation techniques~\cite{li2023dft} or empirical fitting in future work.
\\
To summarize, we have a third approximate constraint required for calculating the functionals $a,b,c$ in eq.~\ref{ansatz_AC}
\begin{align}
    a + \frac{b \sqrt{2}}{c+1} := E^{\rm SCAN}_{\rm x} +  2 E^{\rm SCAN}_{\rm c}
   \label{constraint3}
\end{align}
This completes the construction of our interpolant for the AC, which is on display in figure~\ref{fig:h2_ac} for the H\(_2\) system transitioning from moderate to strongly correlated regime as the bond dissociates.
The corresponding exact (full-CI) AC curves for these systems are available in Ref.~\cite{teale10}.
Evidently, our interpolant captures the correct behaviour across the entire dissociation profile as the two electrons become increasingly correlated.
At the $R=10.0$ Å inter-nuclear separation, the interpolant shows a steep drop to its $W_{\infty}$ limit due to the slope $W'_{0}$ diverging, which causes MP2 and traditional double-hybrid functionals to diverge as well for which regularization schemes have been proposed~\cite{lee2018regularized}.
This divergence is absent within our functional since our AC interpolant has the correct asymptotics (eq.~\ref{asymptote_func}), as evident from figure~\ref{fig:h2_ac} above and the bond-dissociation results below.
\\
The $E_{\rm XC}$ expression of our non-linear and non-empirical (nlane) double-hybrid functional can now be obtained through the AC integration, $E_{\rm xc} = \int_{0}^{1} W_{\lambda}~d\lambda$ :
\begin{align}
   E_{\rm XC}^{\rm nlane} &= E_{\rm X}^{\rm HF} + \frac{4E_{\rm C}^{\rm MP2}}{c(\frac{1}{2} - c)} 
    \left( \sqrt{2} - 1 + \Phi(c) \right) 
    \label{nlane}
\end{align}
where 
\begin{align}
\Phi(c)
= 
\begin{cases}
\sqrt{\frac{1-c}{c}} \left(\tan^{-1}{\sqrt{\frac{c}{1-c}}} - \tan^{-1}{\sqrt{\frac{2c}{1-c}}} ~\right), &  c < 1 \\[1.5em] \frac{1}{2}, & c=1\\[1.5em]
\sqrt{\frac{c-1}{c}} \tanh^{-1}{\frac{\sqrt{c(c-1)}(\sqrt{2} - 1)}{1+ c(\sqrt{2} - 1)}}, & c > 1
\end{cases}
\label{phi_cases}
\end{align}
and $c$ comes from the solution of the three constraint equations,~\ref{constraint1},~\ref{constraint2},~\ref{constraint3}
\begin{align}
     c &=  \frac{\sqrt{9 \alpha^2 - 16\sqrt{2}\alpha + 12\alpha + 4} - \alpha +2}{4 \alpha}, \nonumber 
     \\
     \alpha &= \frac{E_{\rm x}^{\rm SCAN} - E_{\rm x}^{\rm HF} + 2E_{\rm c}^{\rm SCAN}}{2 E_{\rm c}^{MP2}} \label{alpha}
\end{align}
The $\alpha$ ratio in the above equation is very similar to the optimal HF-exchange mixing fraction within global hybrids estimated in Ref.~\cite{burke1997adiabatic} using the two-legged representation of the AC.
We note here that while the function $\Phi(c)$ in eq.~\ref{phi_cases} above is written in a piecewise form, it is continuous and differentiable at $c=1$.
This corresponds to the case when $W_{1}^{\rm SCAN} - E_{\rm x}^{\rm HF} = (4 - 2\sqrt{2})E_{\rm c}^{\rm MP2}$, which is consistent with the monotonicity of the AC~\cite{AC_review_DAAS_Vuckovic}.
The proof for the continuity and differentiability of $\Phi(c)$, and consequently $E_{\rm xc}^{\rm nlane}$, is provided in the supplementary information (SI).
We also note that a single anlytical form of $\Phi(c)$ is not necessary as the numerical integration of eq.~\ref{ansatz_AC} from $\lambda = 0$ to $1$ is quite efficient for arbitrary precision.
Furthermore, the functional derivative $\frac{\delta E_{\rm XC}^{\rm nlane}}{\delta n}$ required for self-consistent calculations can be directly obtained using the derivative of $W_{\lambda}$ in eq.~\ref{ansatz_AC} via the Leibniz rule as also discussed in the SI.
\\
We note here that the expression in eq.~\ref{nlane} is independent of the choice of semi-local DFA for the $W_{1}$ limit in eq.~\ref{w1_scan} which is only present in the $\alpha$ ratio in eq.~\ref{alpha}. 
Due to our use of SCAN in eq.~\ref{w1_scan} in the current work, all our results are denoted as nlane-SCAN.
Hence, eq.~\ref{nlane} fixes the weak correlation limit of SCAN's adiabatic connection through (near) exact values (upto second-order) while retaining the correlated-limit of SCAN.
Furthermore, the interpolation is made using a function (eq.~\ref{ansatz_AC}) well suited for the requirements of the AC (see ref.~\cite{AC_review_DAAS_Vuckovic}).
This is likely the reason why nlane-SCAN improves upon SCAN in virtually all cases as shown in our results below, without using any empirical fitting.
\\
Eq.~\ref{nlane} must be contrasted to traditional, linear (double) hybrid  functionals
\begin{align}
    E^{\rm (double)~hybrid}_{\rm xc} &=  \xi_1 E^{\rm HF}_{\rm x}  + \xi_2 E^{\rm MP2}_{\rm c} + \xi_{3} E^{\rm DFA}_{\rm x} + \xi_{4} E^{\rm DFA}_{\rm c}
    \label{linear_hybrid}
\end{align}
where $\xi_1,\xi_2,\xi_3,\xi_4$ are scaling parameters usually obtained through empirical fitting. 
Note that often $\xi_3 = 1 - \xi_1$ and $\xi_4 = 1-\xi_2$ thereby making $\xi_1$, $\xi_2$ "mixing-fractions" of HF-exchange and MP2 correlation energy respectively.
Eq.~\ref{nlane} does not have such mixing fractions, or any fixed parameters that require empirical fitting.
The HF-exchange and MP2-correlation energies enter through the matching of the Taylor series of our interpolant for the AC (eq.~\ref{constraint1}, eq.~\ref{constraint2}) with exact values from perturbation theory at $\lambda=0$.
Hence, nlane still reduces delocalization errors, as demonstrated through our results below, without the use of such "mixing-fractions".
Furthermore, it is free from one-electron SIE by construction as discussed earlier~\ref{constraint2}.
For comparison, we include the two most popular (empirical) linear hybrid and double hybrid functionals, with forms as in eq.~\ref{linear_hybrid} above, namely B3LYP~\cite{becke93} and B2PLYP~\cite{grimme2007doublehybrid} respectively in all our subsequent results.

\section{Computational methods}
We briefly describe the computational details for our results.
All results reported with nlane-SCAN in this work are calculated non self-consistently, i.e. eq.~\ref{nlane} is evaluated on KS-orbitals supplied from a converged SCF calculation with a different DFA.
For consistency, we use SCAN orbitals throughout but any DFA can be employed for the initial SCF and this choice is provided in our code.
Following a self-consistent SCAN calculation, $E_{\rm x}^{\rm HF}$ and $E_{\rm c}^{\rm MP2}$ (all electron) energies are calculated on these orbitals and the value from eq.~\ref{nlane} is used to replace the SCAN XC-energy.
Our code further provides the option of using any DFA of choice for the $W_{1}$ limit in constraint~\ref{constraint3}, and not just SCAN.

All electron MP2 calculations were used throughout with no frozen orbitals in order to calculate the $E_{\rm c}^{\rm MP2}$ energies used in nlane-SCAN.
No RI-MP2 or density-fitting has been used in our work although the choice for this is also provided in our code.
Unless stated otherwise, the def2-QZVPP basis set~\cite{def2-qzvp} was used in all our calculations.
Other basis sets, as suggested in Ref.~\cite{mehta2022explicitly}, maybe employed due to the slow convergence of MP2 correlation energies with basis-set cardinality.

All open-shell systems were treated within the unrestricted Kohn-Sham framework.
For nlane-SCAN this corresponds to all ingredients in eq.~\ref{alpha} being evaluated twice on the two sets of orbitals and densities for each spin channel.
The final XC-energy is then a summation of the two $E_{\rm XC}$ values evaluated from eq.~\ref{nlane}.
PySCF 2.8.0 package has been used to generate all results~\cite{pyscf}.
See Data and Code section below for the publicly available code and further details.
\\
A major drawback of ACM functionals such as nlane is the lack of size-consistency which linear double hybrids do not suffer from.
The size consistency is, however, straightforward to restore at no additional cost using the correction proposed in Ref.~\cite{vuckovic2018restoring}.
We note that none of our subsequent results with nlane-SCAN have been corrected for the size-consistency error since these were found to be nearly negligible in most cases.
This will be analyzed more thoroughly in a follow-up work alongside  fully self-consistent nlane implementation using the optimized effective potential~\cite{OEP_AQC} based implementation in Ref.~\cite{DHOEPSmiga2016}.

\section{Results and discussion}
\subsection{Atomic total energies}
\begin{table}[h]
    \centering
    \renewcommand{\arraystretch}{1.2}
    \setlength{\tabcolsep}{8pt}
    \begin{adjustbox}{max width=\columnwidth}
    \begin{tabular}{lccccc}
        \toprule
        \textbf{Atom} & \textbf{Exact} & \textbf{SCAN} & \textbf{nlane-SCAN} & \textbf{B3LYP} & \textbf{B2PLYP} \\
        \midrule
        H  & -0.500  & -0.500  & -0.500  & -0.499  & -0.499  \\
        He & -2.904  & -2.905  & -2.904  & -2.908  & -2.904  \\
        Li & -7.478  & -7.480  & -7.474  & -7.482  & -7.474  \\
        Be & -14.667 & -14.650 & -14.657 & -14.659 & -14.655 \\
        B  & -24.654 & -24.641 & -24.646 & -24.647 & -24.643 \\
        C  & -37.845 & -37.839 & -37.841 & -37.839 & -37.837 \\
        N  & -54.589 & -54.589 & -54.588 & -54.580 & -54.581 \\
        O  & -75.067 & -75.072 & -75.066 & -75.069 & -75.066 \\
        F  & -99.734 & -99.745 & -99.733 & -99.739 & -99.736 \\
        Ne & -128.939 & -128.947 & -128.936 & -128.937 & -128.938 \\
        \midrule
        \makecell{\textbf{MAE} \\ \textbf{(kcal/mol)}} & - & 3.95  & \textbf{2.14}  & 3.06  & 3.13 \\
        \bottomrule
    \end{tabular}
    \end{adjustbox}
    \caption{Total atomic energies (in Hartree) for the first 10 chemical elements compared to exact values from Refs.~\cite{gill1992investigation,becke93}.
    Last row shows the mean absolute error (MAE) in kcal/mol.
    All nlane-SCAN results are using SCAN orbitals.}
    \label{tab:total_atomic_energies}
\end{table}
To evaluate the absolute accuracy of the nlane-SCAN functional, we begin with total atomic energies for the first ten elements. 
As described in the preceding section, the current implementation of nlane-SCAN is non-self-consistent and relies on fixed SCAN orbitals and densities. 
While most chemically relevant properties—such as reaction energies and barrier heights—depend only on energy differences, the total energy remains a fundamental quantity in quantum mechanics. For a non-empirical method such as nlane, it is therefore valuable to assess how well it approximates the exact solution of the time-independent Schrödinger equation.

Atoms provide a stringent and well-controlled benchmark for this purpose. They are the basic constituents of all molecules, and for light elements, their exact non-relativistic total energies at the complete basis set (CBS) limit are known from high-level configuration interaction (CI) calculations~\cite{gill1992investigation,becke93}. 
These reference values offer a direct test of the absolute performance of a density functional approximation.

Table~\ref{tab:total_atomic_energies} reports total energies for H to Ne, comparing nlane-SCAN against SCAN, B3LYP, and B2PLYP. 
Despite not being fitted to any atomic data, nlane-SCAN achieves the lowest mean absolute error (MAE) among all four methods, improving upon SCAN in 9 out of 10 cases and outperforming the widely used empirical hybrids B3LYP and B2PLYP. 
This result suggests that nlane-SCAN provides systematically more accurate total energies, and that its improvements in energy differences (shown in the next section) are not merely the result of error cancellation.
\subsection{General thermochemistry}
\begin{figure*}[!ht]
    \centering
    \includegraphics[width=\linewidth]{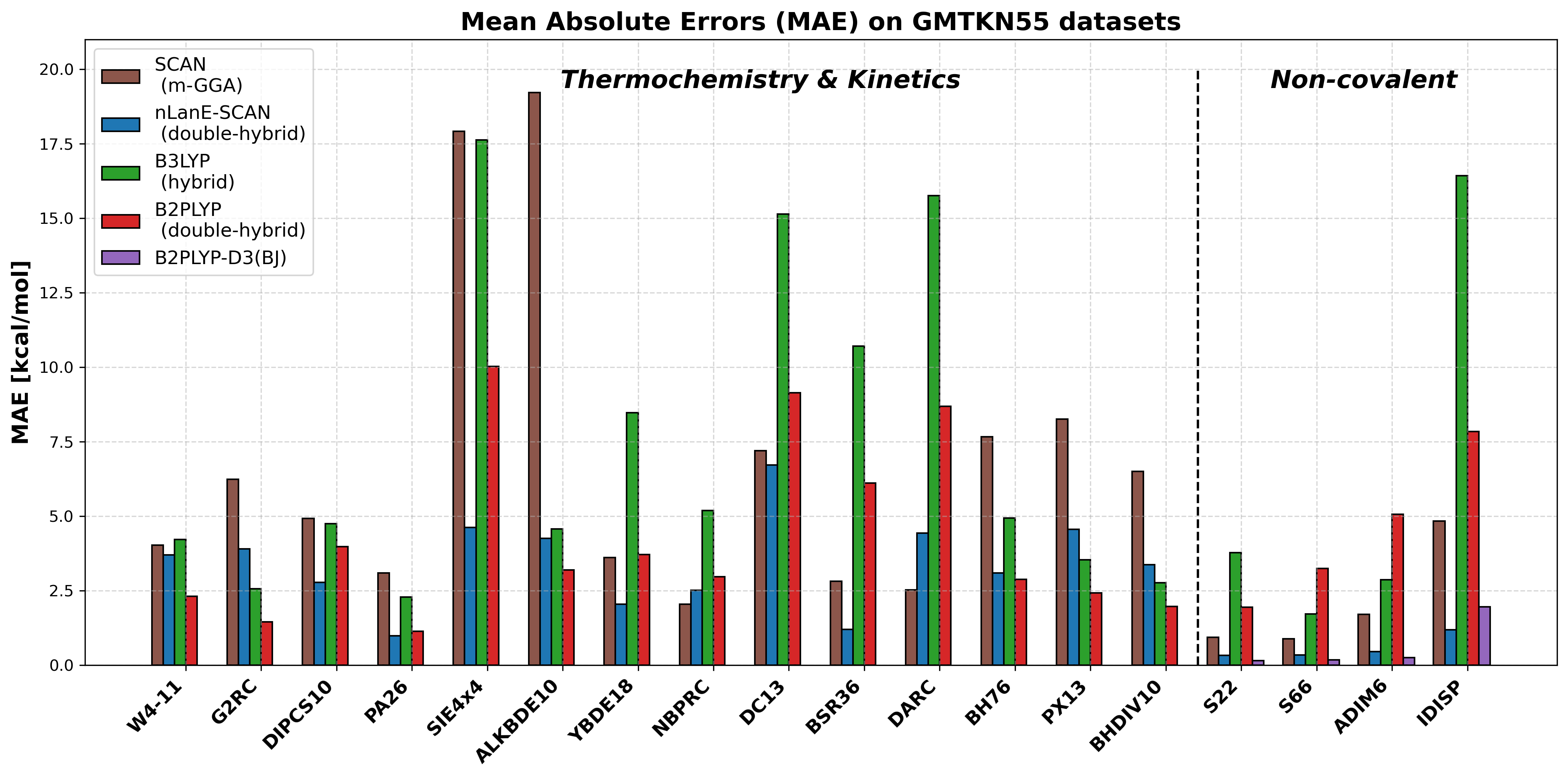}
    \caption{Mean absolute errors (MAE) across 18 datasets of (mostly) small molecules from the GMTKN55 benchmark~\cite{goerigk2017look}.
    Errors for the SCAN, B3LYP, B2PLYP, B2PLYP-D3(BJ) functionals are taken from the GMTKN55 paper.
    All nlane-SCAN results are using SCAN orbitals.
    See Data and code availability section below for code to re-produce these results. See Table 1 in supplementary information for the nlane-SCAN MAE values plotted in this figure.
    }
    \label{fig:gmtkn55}
\end{figure*}
The robustness of nlane-SCAN for general chemical predictions is evaluated using a representative subset of the GMTKN55 benchmark, which spans thermochemistry, kinetics, and non-covalent interactions for main-group molecular systems~\cite{goerigk2017look}.  
Figure~\ref{fig:gmtkn55} presents the mean absolute errors (MAEs) of nlane-SCAN on 18 datasets from GMTKN55 (mostly smaller systems), compared with SCAN, B3LYP, and B2PLYP.  
Unlike B3LYP and B2PLYP---both of which were fitted to molecular energies and reaction barriers from subsets of this benchmark (e.g., G2RC, BH76)---nlane-SCAN contains no empirical parameters and was not trained on any atomic or molecular systems.  
Despite this, it matches or outperforms these empirical functionals across all categories.  
For instance, B3LYP was fitted only to neutral, closed-shell organic molecules and atoms from the G2 dataset~\cite{G2RC-2}, and accordingly performs poorly on chemically distinct systems such as charged, stretched species (SIE4x4~\cite{SIE4x4}), ylides (YBDE18~\cite{YBDE18}), saturated hydricarbon reactions (BSR36~\cite{BSR36-1}), and Diels--Alder cycloadditions (DARC~\cite{DARC-2}).  
In contrast, every dataset in Figure~\ref{fig:gmtkn55} represents an extrapolation test for nlane-SCAN, and yet its performance remains uniformly strong.  
This consistent behavior stems from its non-empirical design, enabling reliable generalization across chemically diverse regions of chemical compound space (CCS).

Notably, nlane-SCAN improves upon SCAN in 16 out of the 18 datasets---sometimes substantially.  
For example, in SIE4x4 and ALKBDE10, the average error is reduced by more than 10 kcal/mol.  
The SIE4x4 benchmark is particularly sensitive to delocalization error, a known deficiency of most DFAs.  
Accurate predictions for these systems typically require a large fraction of exact exchange: Ref.~\cite{bursch2022dispersion} reports that r\textsuperscript{2}SCAN with 50\% HF exchange and empirical dispersion correction yields an MAE of 4.6 kcal/mol.  
However, this level of exchange is not optimal across all systems---25\% was found to be optimal for all other tasks in the same study.  
nlane-SCAN, by contrast, achieves a similar error of 4.62 kcal/mol without mixing fractions or empirical correction, and performs significantly better than both B3LYP and B2PLYP on this benchmark.  
Its accuracy emerges from the exact constraints in eq.~\ref{constraint1},~\ref{constraint2} and physically motivated non-linear mixing, avoiding the need to tune arbitrary fractions of exchange or correlation.  
Across the full benchmark, it also consistently outperforms B2PLYP except in barrier height datasets (e.g., BH76, PX13, BHDIV10), where the SCAN approximation to \(W_1\) may be insufficiently accurate.

Finally, nlane-SCAN also improves upon SCAN in non-covalent interaction datasets, which probe dispersion and long-range correlation.  
Although SCAN is capable of capturing mid-range dispersion through its semi-local construction, nlane-SCAN outperforms it as well as the non-local hybrids B3LYP and B2PLYP.  
Its performance is competitive even with B2PLYP-D3(BJ)~\cite{d3-bj}, which includes empirical dispersion corrections.  
Similar accuracy was previously observed for ISI-type ACMs~\cite{isiHF_analysis2016}, where the inclusion of the \(W'_\infty\) functional enables the description of long-range coupled electron oscillations. 
Although nlane-SCAN does not explicitly include \(W_\infty\) or \(W'_\infty\), we attribute its ability to capture dispersion to the approximate resummation of the divergent GL perturbation expansion around the weak-coupling limit in Eq.~\ref{GLn}.  
While some error cancellation between overestimated (MP2-like) and underestimated (SCAN) dispersion energies may contribute, this effect is likely small since nlane does not mix these ingredients linearly (see Eq.~\ref{nlane}).  
A more detailed analysis of the origin of dispersion accuracy in nlane-SCAN will be explored in future work.
\subsection{Bond dissociations}
\begin{figure*}[!ht]
    \centering
    \includegraphics[width=0.7\linewidth]{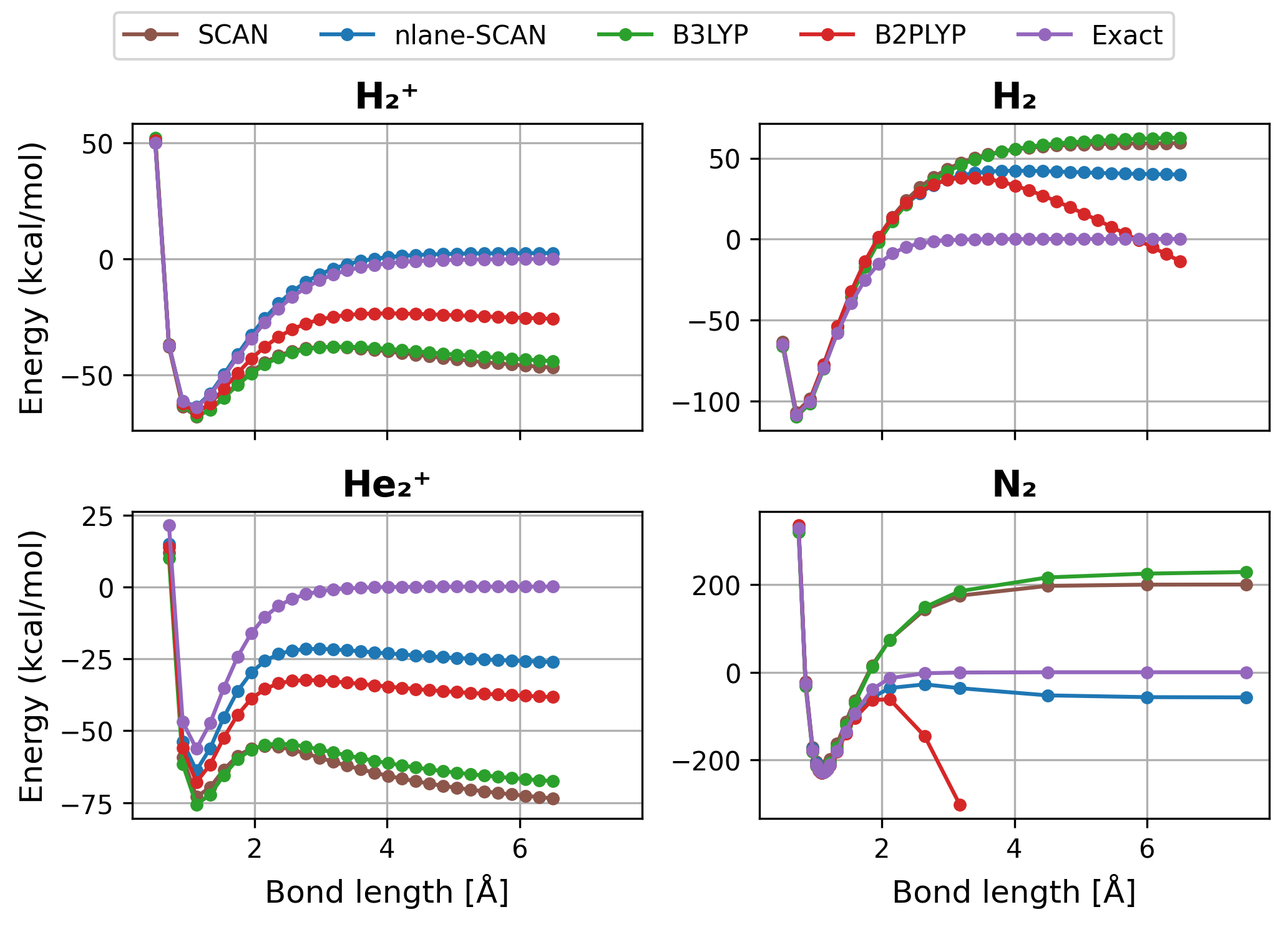}
    \caption{Bond dissociation curves for four diatomic molecules. 
    Unrestricted calculations were used for the open-shell species (H\(_2^+\) and He\(_2^+\)), while restricted calculations were performed for H\(_2\) and N\(_2\).  
    The "Exact" reference corresponds to all-electron CCSD(T) results for all systems except N\(_2\), where we use the accurate $r_{12}$-MR-ACPF values reported in Table 2 of Ref.~\cite{N2_ref1998}.  
    All DFT and CCSD(T) calculations employed the cc-pVQZ basis set~\cite{dunning:1989:bas}.
    All nlane-SCAN results are using SCAN orbitals.}
    \label{fig:bond_dissoc}
\end{figure*}
Figure~\ref{fig:bond_dissoc} presents bond dissociation energy curves for two charged (unrestricted Kohn--Sham) and two neutral (restricted Kohn--Sham) diatomic molecules.  
Across all four systems, nlane-SCAN consistently improves upon SCAN and outperforms the B3LYP and B2PLYP functionals, yielding the most accurate results throughout the entire dissociation range.

In the case of H\(_2^+\), nlane-SCAN nearly exactly reproduces the correct dissociation behavior.  
This is expected since it is free from one-electron SIE as discussed earlier in eq.~\ref{constraint2}.  
For the multi-electron He\(_2^+\) system as well, nlane-SCAN shows significant improvement over SCAN, demonstrating reduced delocalization errors in general.  
This arises from the incorporation of two exact constraints in the weak-coupling limit---namely, the correct linear coefficient (Eq.~\ref{constraint1}) and the accurate initial slope of the AC integrand (Eq.~\ref{constraint2})---which together enforce a more physically realistic description of fractional charge dissociation.

In the strongly correlated, closed-shell cases of H\(_2\) and N\(_2\), the HOMO--LUMO gap vanishes upon bond stretching, leading to a divergence in the MP2 correlation energy (Eq.~\ref{Ec_mp2}) and, consequently, in the B2PLYP double hybrid functional which linearly mixes MP2 with DFA correlation.  
Although nlane-SCAN also employs MP2 correlation, it does so only to evaluate the initial slope of the adiabatic connection (Eq.~\ref{constraint2}).  
The total correlation energy is instead modeled through a non-linear functional form (Eq.~\ref{ansatz_AC}) that incorporates the correct strong-coupling asymptotic expansion (Eq.~\ref{asymptote_func}).  
As a result, nlane-SCAN avoids divergence and exhibits stable and accurate behavior even in these multi-reference, strongly correlated regimes, improving upon SCAN throughout dissociation.

We anticipate that further improvements are possible with a future self-consistent implementation of nlane, which would provide more accurate orbitals and densities across all systems.

\section{Conclusion}
In this chapter, we have introduced a non-linear and non-empirical double hybrid density functional, termed nlane, constructed through an accurate modeling of the adiabatic connection (AC).  
While nlane shares conceptual similarities with adiabatic connection models (ACMs) such as ISI and SPL, it departs from their traditional construction by replacing the explicit constraint from the strong-coupling limit (\(\lambda \to \infty\)) with the physical, fully interacting limit (\(\lambda = 1\)) as computed from a highly accurate density functional---in this case, SCAN.  
This key design choice allows nlane-SCAN to inherit the non-empirical character and rigorous constraint satisfaction of wavefunction-based approaches from the weak-coupling limit, while retaining, and even improving upon, the accuracy of SCAN for covalently bonded systems in equilibrium.
Furthermore, nlane remains stable for strongly correlated systems by incorporating the correct asymptotic expansion of the AC.
Consequently, this approach reduces delocalization errors while simultaneously improving the description of strong static correlation effects through a physically motivated combination of wavefunction and semi-local density functional theory.

Unlike conventional double hybrids, which often rely on empirical mixing parameters and linear interpolation of perturbative correlation energies, nlane-SCAN employs a non-linear, physically motivated interpolation ansatz that enforces the correct asymptotic behavior of the AC integrand.  
By correcting SCAN’s weak-coupling behavior up to second order while preserving its accurate \(W_1\) energy at full interaction strength, nlane-SCAN achieves systematic improvements across a wide range of systems.  
As demonstrated in Table~\ref{tab:total_atomic_energies} and Figure~\ref{fig:gmtkn55}, it delivers lower mean absolute errors than SCAN, B3LYP, and even B2PLYP for total atomic energies and general thermochemical benchmarks, despite having no fitted parameters.  
This consistent accuracy highlights the transferability and extrapolative power of the nlane framework, distinguishing it from empirically tuned hybrid and double hybrid functionals.

Moreover, nlane-SCAN inherits many of the formal advantages of ACM-based functionals.  
In particular, it remains well-behaved for challenging bond dissociation problems where conventional double hybrids fail due to divergence in the MP2 correlation energy (Figure~\ref{fig:bond_dissoc}).  
Since MP2 enters nlane only through the initial slope of the AC integrand, rather than as a linearly mixed component, the energy remains finite and accurate even in the presence of near-degeneracy and strong correlation.  
The improved dissociation behavior observed for H\(_2\), H\(_2^+\), He\(_2^+\), and N\(_2\) suggests that the interpolating function used in nlane captures essential features of both weak and strong correlation without needing an explicit treatment of the \(\lambda \to \infty\) limit.  
This balance between physical accuracy and numerical stability makes nlane-SCAN a promising candidate for general-purpose quantum chemical applications.

Future work will extend this framework in multiple directions.  
One natural next step is to incorporate both the exact strong-coupling constraint and the fully interacting limit into a unified interpolating scheme, potentially combining the strengths of ISI-type ACMs with the practical benefits of the current approach.  
We also aim to systematically analyze the impact of using different parent DFAs for \(W_1\) and their compatibility with the nlane formalism.  
In addition, the current implementation is non-self-consistent, and future developments will explore the benefits of a fully self-consistent nlane calculation using optimized orbitals.  
Such a formulation is expected to further improve accuracy in both energy predictions and electronic properties, while also allowing for an assessment of size-consistency errors, which are known to affect non-linear double hybrids.  
These investigations will deepen our understanding of physically motivated interpolation in DFT and further expand the utility of AC-based functionals in real-world quantum chemical simulations.

\section*{Data and Code}
\label{data_code}
See \verb|github.com/dkhan42/nlane-DH| for PySCF~\cite{pyscf} based code for performing nlane calculations with any choice of DFA for the SCF and $W_{1}$ limit. 
A density-fitting based implementation is also available.

\section*{Acknowledgements}
I am highly grateful to Prof. Szymon Śmiga and my supervisor Prof. Anatole von Lilienfeld for their valuable advice, insight and feedback on this work. 
I thank Simon Krug for insightful discussions.
I acknowledge support from the Vector Institute in form of a graduate research fellowship, and the University of Toronto for resources and funding.

\bibliography{main}

\end{document}


\title{Supplementary information}
\date{}
\maketitle
\section{Proof that $E_{\rm XC}^{\rm nlane}$ is continuous and differentiable} \label{sec:proof_differentiable_nlane}

From the piecewise definition of the function $\Phi(c)$, it is obvious that $E_{\rm XC}^{\rm nlane}$ is continuous and differentiable for the cases $c<1$ and $c>1$.
Herein we provide the proof, alongside the limit, for the special case when $c=1$.
This corresponds to the ratio $\alpha$ being equal to $2 - \sqrt{2}$ implying that $W_{1}^{\rm SCAN} - E_{\rm X}^{\rm HF} = (4 - 2\sqrt{2})E_{\rm C}^{\rm MP2}$.
We first define the function $f(c)$, which is equivalent to $E_{\rm XC}^{\rm nlane}(c)$ through the definition of the AC and the $W_{\lambda}$ expression used in nlane:
\begin{align}
    E_{\rm XC}^{\rm nlane}(c) = f(c) = \int_0^1 \left( a + \frac{b \sqrt{x+1}}{c x + 1} \right) dx, 
\end{align}
herein $a, b \in \mathbb{R}$ are constants.
We need to show that $f(c)$ is continuous and differentiable at $c=1$.

\subsection{Continuity at \( c = 1 \)}

We must show that for every \( \varepsilon > 0 \), there exists \( \delta > 0 \) such that
\begin{align}
|c - 1| < \delta \quad \Rightarrow \quad |f(c) - f(1)| < \varepsilon.
\end{align}

Note that
\begin{align*}
|f(c) - f(1)|
&= \left| \int_0^1 \left( a + \frac{b \sqrt{x+1}}{c x + 1} \right) dx - \int_0^1 \left( a + \frac{b \sqrt{x+1}}{x + 1} \right) dx \right| \\
&= \left| \int_0^1 b \sqrt{x+1} \left( \frac{1}{c x + 1} - \frac{1}{x + 1} \right) dx \right| \\
&\leq |b| \int_0^1 \sqrt{x+1} \left| \frac{1}{c x + 1} - \frac{1}{x + 1} \right| dx.
\end{align*}

Now, we estimate the difference of the reciprocals:
\begin{align}
\left| \frac{1}{c x + 1} - \frac{1}{x + 1} \right| = \left| \frac{(1 - c)x}{(c x + 1)(x + 1)} \right| \leq |1 - c| \cdot \frac{x}{(c x + 1)(x + 1)}.
\end{align}

We restrict \( c \in [1 - \delta_0, 1 + \delta_0] \) for some \( \delta_0 \in (0, 1) \), so that \( c x + 1 \geq 1 - \delta_0 > 0 \) for all \( x \in [0,1] \). 
Then:
\begin{align}
\left| \frac{1}{c x + 1} - \frac{1}{x + 1} \right| \leq |1 - c| \cdot \frac{x}{(1 - \delta_0)(x + 1)}.
\end{align}

Also note that \( \sqrt{x+1} \leq \sqrt{2} \) and \( x \leq 1 \). 
Hence :
\begin{align}
|f(c) - f(1)| \leq |b| \cdot |1 - c| \cdot \int_0^1 \frac{x \sqrt{x+1}}{(1 - \delta_0)(x + 1)} dx \leq \frac{|b| \sqrt{2} |1 - c|}{1 - \delta_0} \int_0^1 x \, dx = \frac{|b| \sqrt{2} |1 - c|}{1 - \delta_0} \cdot \frac{1}{2}.
\end{align}

Now, fix \( \delta_0 = \frac{1}{2} \), implying :
\begin{align}
|f(c) - f(1)| \leq |1 - c| \cdot \frac{|b| \sqrt{2}}{1 - \delta_0} \cdot \frac{1}{2} = |1 - c| \cdot |b| \sqrt{2}.
\end{align}

To make this less than \( \varepsilon \), it suffices to take:
\begin{align}
|1 - c| < \frac{\varepsilon}{|b| \sqrt{2}}.
\end{align}

Let \( \varepsilon > 0 \) be given. 
Set \( \delta = \frac{\varepsilon}{|b| \sqrt{2}} \). 
Then for all \( c \in \mathbb{R} \) such that \( |c - 1| < \delta \), we have:
\begin{align}
|f(c) - f(1)| < \varepsilon.
\end{align}

\noindent Hence, \( f \) is continuous at \( c = 1 \) with $f(1) = a + 2(\sqrt{2} - 1)b$

\subsection{Differentiability at \( c = 1 \)}

We now prove that \( f \) is differentiable at \( c = 1 \) and compute \( f'(c) \). 
First, compute the partial derivative of the integrand with respect to \( c \):
\begin{align}
\frac{\partial}{\partial c} \left( \frac{b \sqrt{x+1}}{c x + 1} \right) = -b \cdot \frac{x \sqrt{x+1}}{(c x + 1)^2}.
\end{align}

This partial derivative is continuous in both \( x \) and \( c \) on the domain where \( c x + 1 > 0 \). Furthermore, for \( c \) in a neighborhood of 1 and \( x \in [0,1] \), we have
\begin{align}
\left| \frac{\partial}{\partial c} \left( \frac{b \sqrt{x+1}}{c x + 1} \right) \right| = \left| b \cdot \frac{x \sqrt{x+1}}{(c x + 1)^2} \right| \leq |b| \cdot \frac{x \sqrt{2}}{1^2} = |b| \sqrt{2} x.
\end{align}
The function \( x \mapsto |b| \sqrt{2} x \) is integrable over \( [0,1] \).

Now, by the Leibniz rule we have:
\begin{align}
f'(c) = \frac{d}{dc} \int_0^1 \left( a + \frac{b \sqrt{x+1}}{c x + 1} \right) dx = \int_0^1 \frac{\partial}{\partial c} \left( \frac{b \sqrt{x+1}}{c x + 1} \right) dx = -b \int_0^1 \frac{x \sqrt{x+1}}{(c x + 1)^2} dx
\label{leibniz_derivative_nlane}
\end{align}

This expression defines a continuous function of \( c \), since the integrand is finite when $cx \geq 0$ as is the case herein.

Hence, $f$ is differentiable at $c=1$ with
\begin{align}
f'(1) = - b \int_0^1 \frac{x \sqrt{x+1}}{(x + 1)^2} dx \approx -0.24264 b
\end{align}
via numerical integration.
Eq.~\ref{leibniz_derivative_nlane} allows calculating functional derivatives of nlane without requiring an analytical form of the corresponding $E_{\rm XC}^{\rm nlane}$.
This approach holds for all ACMs.
\section{GMTKN55 mean-absolute errors}
\begin{table}[!h]
    \centering
    \renewcommand{\arraystretch}{1.2}
    \setlength{\tabcolsep}{8pt}
    \begin{tabular}{lc}
        \toprule
        \textbf{Dataset} & \textbf{MAE [kcal/mol]} \\
        \midrule
        W4-11     & 3.70  \\
        G2RC      & 3.90  \\
        DIPCS10   & 2.78  \\
        PA26      & 0.98  \\
        SIE4x4    & 4.62  \\
        ALKBDE10  & 4.25  \\
        YBDE18    & 2.04  \\
        NBPRC     & 2.51  \\
        DC13      & 6.72  \\
        BSR36     & 1.20  \\
        DARC      & 4.43  \\
        BH76      & 3.09  \\
        PX13      & 4.56  \\
        BHDIV10   & 3.37  \\
        S22       & 0.33  \\
        S66       & 0.34  \\
        ADIM6     & 0.45  \\
        IDISP     & 1.19  \\
        \bottomrule
    \end{tabular}
    \caption{nlane-SCAN mean absolute errors (MAE) on the GMTKN55 datasets plotted in Figure 2 of main text.}
    \label{tab:nlane_errors}
\end{table}